**Raman Characterization of Platinum Diselenide Thin Films**

Maria O'Brien[a,b†], Niall McEvoy[a,b†*], Carlo Motta[a,c], Jian-Yao Zheng[a,c], Nina C. Berner[a,b], Jani Kotakoski[d], Kenan Elibol[d], Timothy J. Pennycook[d], Jannik C. Meyer[d], Chanyoung Yim[a,b], Mohamed Abid[e], Toby Hallam[a,c], John F. Donegan[a,c], Stefano Sanvito[a,c] and Georg S. Duesberg[a,b*]

[a] Centre for the Research on Adaptive Nanostructures and Nanodevices (CRANN) and Advanced Materials and BioEngineering Research Institute (AMBER), Trinity College Dublin, Dublin 2, Dublin, Ireland

[b] School of Chemistry, Trinity College Dublin, Dublin 2, Dublin, Ireland

[c] School of Physics, Trinity College Dublin, Dublin 2, Dublin, Ireland

[d] Faculty of Physics, University of Vienna, Boltzmanngasse 5, A-1090 Vienna, Austria

[e] KSU-Aramco Center, King Saud University, Riyadh 11451, Saudi Arabia

†These authors contributed equally.

*E-mail: nmcevoy@tcd.ie and duesberg@tcd.ie

Platinum diselenide ($PtSe_2$) is a newly discovered 2D material which is of great interest for applications in electronics and catalysis. $PtSe_2$ films were synthesized by thermally-assisted selenization of predeposited platinum films and scanning transmission electron microscopy revealed the crystal structure of these films to be 1T. Raman scattering of these films was studied as a function of film thickness, laser wavelength and laser polarization. $E_g$ and $A_{1g}$ Raman active modes were identified using polarization measurements in the Raman setup. These modes were found to display a clear position and intensity dependence with film thickness, for multiple excitation wavelengths, and their peak positions agree with simulated phonon dispersion curves for PtSe$_2$. These results highlight the practicality of using Raman spectroscopy as a prime characterization technique for newly-synthesized 2D materials.

**Introduction**

2D materials have recently been in the spotlight of the research community due to their fascinating optical and electronic properties[1-5]. While much initial work focused on graphene, a burgeoning interest in other layered materials has developed. This definition of layered materials encompasses the heavily-studied transition metal dichalcogenides (TMDs), MoS$_2$, MoSe$_2$, WS$_2$, and WSe$_2$, but also includes more exotic (and to date less-studied) materials such as black phosphorous, transition metal oxides and hexagonal boron nitride. These materials have in common a layered structure, which can exist from a single monolayer less than a nanometer thick, to a bulk crystal consisting of millions of layers held together by weak van der Waals forces. In MoS$_2$, for example, these single atomic layers typically consist of two hexagonal planes of sulfur atoms arranged around an interstitial hexagonal plane of Mo atoms in a trigonal prismatic, or 2H, arrangement. This geometry and its resulting symmetry leads to several interesting properties, such as an increased direct bandgap in the monolayer[1], and an evolution of Raman spectra with increasing layer thickness[6]. Other layered materials exhibit similar trends with changing thickness, which can potentially be harnessed for emerging applications in electronics and optoelectronics.

Raman spectroscopy is a powerful and non-destructive characterization technique which is extensively used for 2D materials including graphene[7] and TMDs[8]. Herein we focus on the Raman characterization of a new TMD, PtSe$_2$. Initial reports have emerged on the epitaxial growth of PtSe$_2$ on a highly-crystalline Pt substrate under ultra-high vacuum (UHV) conditions[9]. This study identified a transition from semi-metallic to semiconducting behavior with a reduction in layer number to bi- or monolayer, suggesting potential for an array of applications in electronics. Additionally, $PtSe_2$ has previously been synthesized by chemical means for use in graphene hybrid materials[10] and has been theoretically predicted to be a high-performance photo-catalyst[11]. While Raman spectroscopy is used extensively for the characterization of TMDs and other nanomaterials, the Raman characteristics of $PtSe_2$ have not previously been reported. Here we present synthesis of $PtSe_2$ films of controllable thickness by direct selenization of predeposited Pt precursor layers and study the evolution of Raman spectra with thickness. We confirm the stoichiometry of the films to be $PtSe_2$ using X-ray photoelectron spectroscopy (XPS) and the crystal structure to be 1T with scanning transmission electron microscopy (STEM). We identify the in-plane and out-of-plane Raman-active vibrations in $PtSe_2$, and investigate the frequency and intensity dependence of each Raman mode on thickness. The trends in the Raman peak shift agree with our calculated phonon dispersion curves for varying thicknesses of $PtSe_2$.

**Experimental Analysis**

**Synthesis, Materials and Methods**

$PtSe_2$ thin films were fabricated using a similar process to that previously described for other TMDs.[12-14] Briefly, Pt thin films of varying thicknesses were sputtered from a MaTeck Pt target using a Gatan Precision Etching and Coating System (PECS) onto silicon (Si) substrates with 300 nm dry thermal oxide ($SiO_2$). The deposition rate and film thickness were monitored with a quartz-crystal microbalance. These films were then placed in the center of a quartz-tube furnace and heated to a growth temperature of 450 °C under 150 sccm of 10% $H_2$/Ar flow. Selenium (Se) vapor was then produced by heating Se powder (Sigma Aldrich, ≥ 99.99%) to ~220 °C in an independently-controlled, upstream heating zone of the furnace, and carried downstream to the Pt films for a duration of 2 hours, after which the furnace was cooled to room temperature. A schematic of the furnace is shown in the supplementary data, Figure S.1. Henceforth, film thicknesses will be referred to by their starting Pt thickness. Raman spectroscopy was performed using a WITec Alpha 300R with 532 nm and 633 nm excitation lasers with a spectral grating of 1800 lines/mm and a 100× microscope objective (0.95 N.A., spot size ~0.3 μm). Spectra were taken with a laser power of < 300 μW in order to minimize sample heating. Polarized Raman measurements were conducted with a confocal microscope (Renishaw InVia System). A 100× (0.7 NA) microscope objective was used to focus the laser beam and collect the scattered light. An excitation line of 488 nm with a spot size of ~ 0.7 μm was used. Density functional theory calculations were performed with the all-electron code FHI-aims[15] and the generalized gradient approximation (GGA). The "tight" basis set was employed, and an 8×8 k-points grid was used to sample the in-plane Brillouin zone. All the geometries were optimized with a force tolerance of 0.001 eV/Å. Long-range van der Waals interactions were included with the Tkatchenko-Scheffler scheme[16]. Phonon spectra were computed with the Phonopy code[17] with finite displacements of 0.001 Å. XPS spectra were recorded under UHV conditions (<$10^{-8}$ mbar) on a VG Scientific ESCAlab MkII system using Al K$_\alpha$ X-rays and an analyzer pass energy of 20 eV. After subtraction of a Shirley background, the core-level spectra were fitted with Gaussian-Lorentzian and Doniach-Sunjic (for the metallic Pt 4f component) line shapes using the software CasaXPS. For the STEM analysis, an as-grown $PtSe_2$ thin film of 0.5 nm starting Pt thickness was transferred onto a holey carbon TEM grid. A polymer support technique was employed for the transfer process, whereby polymethyl methacrylate (PMMA, MicroChem) was spin-coated onto the as-grown sample and baked in atmosphere at 150 °C for 5 min. The films were then floated on 2 M NaOH at 80 °C until the $PtSe_2$/PMMA films were left floating on the surface. After cleaning in deionized water the films were transferred onto grids and allowed to dry in a desiccator. The PMMA support layer was then dissolved in acetone and the films baked prior to STEM measurements. Atomic-resolution images were obtained with the scanning transmission electron microscope, Nion UltraSTEM 100, using a high angle annular dark field (HAADF) detector. The microscope was operated at 60 kV.

**Structural and Chemical Characterization of $PtSe_2$**

In Figure 1(a), HAADF STEM images of $PtSe_2$ grown from 0.5 nm Pt are presented in order to investigate the crystal structure of suspended $PtSe_2$, free of substrate interactions or epitaxial growth influences, which has not previously been reported. Previous studies have revealed $PtSe_2$ to have an octahedral, or 1T, type crystal structure[9], more commonly known as the $CdI_2$ crystal group, which also encompasses other TMDs such as $HfS_2$ and $SnS_2$. The results presented here show regions of different thicknesses and crystal orientations, similar to previous reports on the production of TMDs by vapor-phase sulfurization[18-20]. It was possible to identify monolayers and bilayers of $PtSe_2$ in the $CdI_2$ crystal structure in Figure 1(a), which is shown in the schematic in Figure 1(b) and the overlay of theoretical and experimental structure in Figure 1(c). Image simulations of monolayer and bilayer $PtSe_2$ performed with QSTEM software[21] are presented in Figure S.2 in the supplementary data, which confirm the interpretation of layer thicknesses. XPS spectra were recorded on a $PtSe_2$ thin film synthesized from a starting Pt film thickness of 0.5 nm. The Pt 4f core-level spectrum for each film shows primarily peaks associated to $PtSe_2$. The Pt 4f can be deconvoluted into three contributions, the main one at ~72.3 eV attributed to $PtSe_2$ and two smaller ones at ~74 eV and ~71 eV to oxides and unreacted Pt metal, respectively. The relative atomic percentages are shown in Figure 1(d). The Pt $5p_{3/2}$ peak also lies in the same region, but has not been used for analysis. The Se 3d peak is shown in the supplementary data, Figure S.3, and can be deconvoluted into two contributions, one from $PtSe_2$ and another one at higher binding energies from edge or amorphous Se.

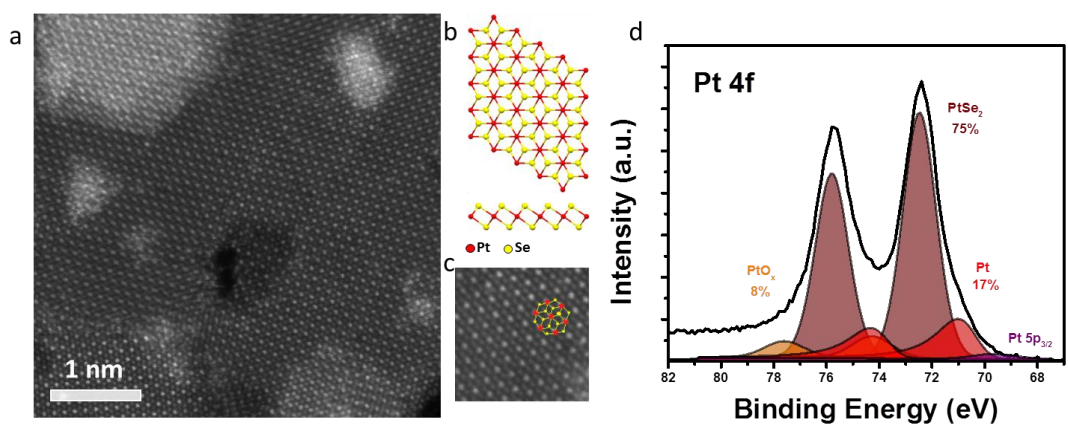

Figure 1 – (a) HAADF STEM of $PtSe_2$ displaying regions of varying layer number (b) Crystal structure and (c) crystal structure overlay of $PtSe_2$ on HRTEM monolayer (d) XPS spectrum of 0.5 nm $PtSe_2$ film.

**Raman Spectroscopy of $PtSe_2$**

Raman spectroscopy is a prime technique for the study of nanomaterials, due to its ease of use and non-destructive characterization. The Raman spectra of TMDs are generally characterized by two main peaks corresponding to the in-plane and out-of-plane motions of atoms. The CdI$_2$ lattice type configuration of PtSe$_2$ results in a $D_{3d}$ point-group symmetry with 3 atoms in a unit cell. The $D_{3d}$ symmetry results in the following irreducible representations of modes at the center of the Brillouin zone:

$$\Gamma = A_{1g} + E_g + 2A_{2u} + 2E_u$$

The two Raman-active modes are labelled as $A_{1g}$ and $E_g$, while the infrared (IR) active modes are $A_{2u}$ and $E_u$. Here we focus on the Raman active modes. In Figure 2(a), the calculated phonon-dispersion curves of monolayer PtSe$_2$ are presented with the corresponding density of states (DOS) shown beside them. A schematic of the vibrational modes in PtSe$_2$ is shown in Figure 2(b), with arrows drawn as guides to show the origin of each mode on the phonon dispersion curve. These include the in-plane $E_g$ mode, describing the in-plane vibration of selenium atoms in opposite directions within a single layer, and the $A_{1g}$ mode describing the out-of-plane vibration of selenium atoms. The calculated phonon dispersion curve allows a theoretical prediction to be made about the frequencies of both Raman and infrared (IR) active vibrations. It should be noted that GGA usually underestimates the phonon frequencies as a result of a tendency to overestimate the equilibrium volume[22]. Nevertheless, as shown in the results in Table S.1 in the supplementary data, it was possible to calculate phonon frequencies for all Raman and IR vibrational modes as predicted by symmetry considerations. The electronic bandstructures of monolayer and bilayer PtSe$_2$ have also been calculated, and are available in the supplementary data, Figure S.4. For the monolayer, the band gap was found to be indirect, with a magnitude of ~1.6 eV. This decreases significantly in the bilayer, where the gap remains indirect but reduces to ~0.8 eV. The band-gap transitions are indicated by black arrows in Figure S.4. Bulk PtSe$_2$ instead is shown to have a semi-metallic band structure, in agreement with previous predictions[9]. The calculations presented here therefore agree with the trend of decreasing the bandgap as the number of layers gets larger. Such radical change in the band structure upon changing the number of layers is common in 2D materials, due to changes in the dielectric screening of long-range Coulombic forces between layers as the layer number decreases[22]. The emerging bandgap in PtSe$_2$ thin layers suggest that the material may be of high interest for future electronic and optoelectronic applications. As a further check, the electronic bandstructure obtained by including spin-orbit coupling (SOC) has also been computed, as shown in Figure S.5 of the supplementary data. The results suggest that SOC has virtually no effect on the electronic properties, especially in the optical region.

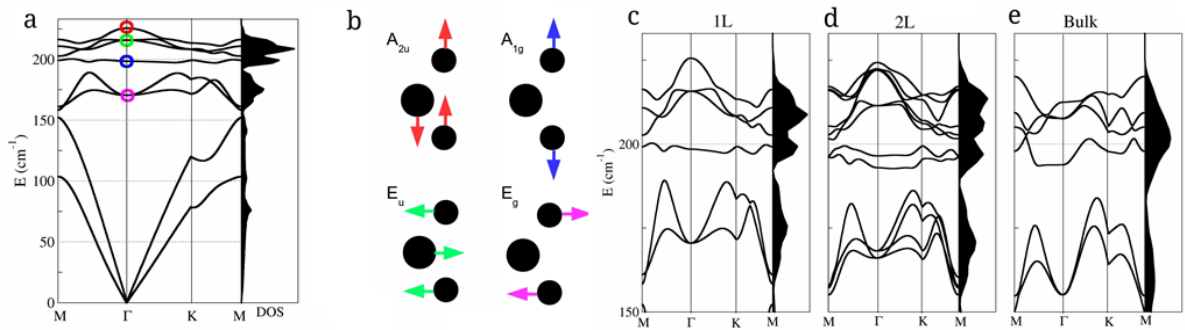

Figure 2 – (a) Calculated phonon dispersion and DOS of monolayer PtSe$_2$. (b) Schematic diagrams of vibrational modes in PtSe$_2$ layers. Large spheres are Se atoms; smaller spheres are Pt atoms. The colors associate each mode with its energy in the bandstructure. (c, d, e) Calculated phonon dispersion curves of monolayer, bilayer and bulk PtSe$_2$.

Experimental Raman spectroscopy measurements were taken on as-synthesized PtSe$_2$ thin films. In Figure 3(a) and (b), Raman spectra of PtSe$_2$ layers from starting Pt thicknesses of 0.5 and 5 nm respectively are shown for 488, 532 and 633 nm excitation wavelengths. Two prominent vibrational modes are visible, at ~175 and ~205 cm$^{-1}$, with a less intense contribution at ~230 cm$^{-1}$. By comparison with the phonon frequency predictions for perfect crystals in Figure 2, the peaks at 175 and 205 cm$^{-1}$ are assigned to $E_g$ and $A_{1g}$ modes, respectively. The differences between theoretical and experimental Raman frequencies are attributed to previously discussed GGA estimation errors and the fact that the thin films studied here do not consist solely of pristine layers of definite thicknesses. The less prominent feature at ~230 cm$^{-1}$ is assigned to an overlap between the $A_{2u}$ and $E_u$ modes, which are longitudinal optical (LO) modes involving the out-of-plane and in-plane motions of platinum and selenium atoms respectively. The relative intensity of this contribution is strongest for the 0.5 nm film where it is evident that it stems from two separate peaks at ~230 cm$^{-1}$ and ~237 cm$^{-1}$. These modes are much less intense for the 5 nm film and manifest as a broad feature centered at ~ 230 cm$^{-1}$. A schematic of the $A_{2u}$ and $E_u$ mode vibrations is shown in Figure 2(b).

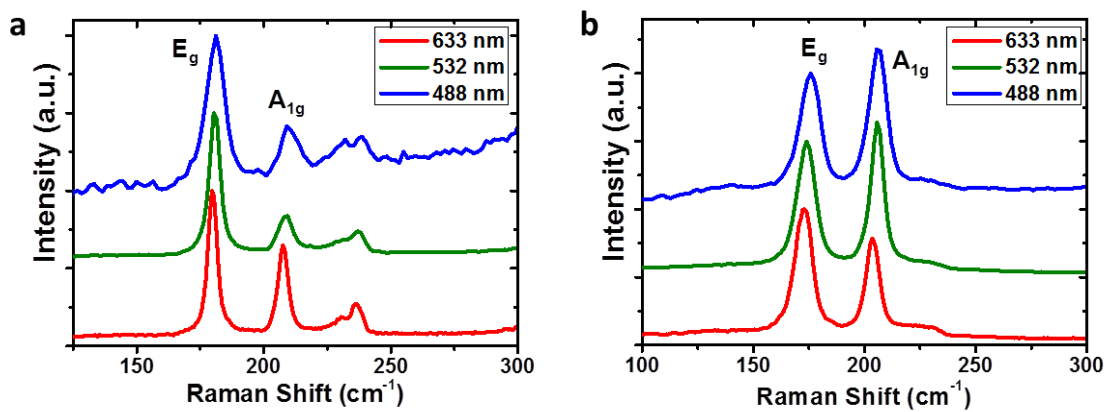

Figure 3 – Raman spectra of PtSe$_2$ films taken using blue (488 nm), green (532 nm) and red (633 nm) excitation wavelengths. (a) 0.5 nm thickness spectra (b) 5 nm thickness spectra. These spectra have been normalized to the $E_g$ mode at ~175 cm$^{-1}$ for clarity.

Polarization-dependent Raman studies, using a 488 nm excitation laser, were carried out to support the peak assignment detailed above. Porto's notation *A(BC)D* is used to describe the Raman geometry and polarization, where A and D represent the wavevector direction of the incoming and the scattered light, respectively, while B and C represent the polarization direction of the incoming and scattered light[23]. In our experiment, [100], [010] and [001] axes are labelled X, Y and Z respectively. Figure 4(a) shows spectra of a 1 nm $PtSe_2$ thin film. For the *Z(YY)$\bar{Z}$* polarization, both $E_g$ and $A_{1g}$ peaks can be observed. When in the *Z(YX)$\bar{Z}$* polarization, there is a clear decrease in the intensity of the $A_{1g}$ mode, confirming the out-of-plane nature of this mode. In Figure 4(b) spectra of a 5 nm $PtSe_2$ thin film are shown. Similar to the 1 nm film, when in the *Z(YY)$\bar{Z}$* polarization, both $E_g$ and $A_{1g}$ peaks can be observed. Note that the $A_{1g}$ peak is larger in the 5 nm film due to increased out-of-plane interactions between more layers in thicker samples. For the *Z(YX)$\bar{Z}$* configuration, there is a clear decrease in the intensity of the $A_{1g}$ mode, further confirming the out-of-plane behavior of this mode, but it does not disappear completely. This can be attributed to the polycrystalline nature of the sample, as shown in the STEM analysis. While the material synthesized is highly crystalline, the crystal grains are small (with lateral dimensions on the order of a few nm) and randomly oriented, which leads to $A_{1g}$ contributions that are diminished but not entirely absent in the thicker samples, where more edge-on orientation of films could be present[24].

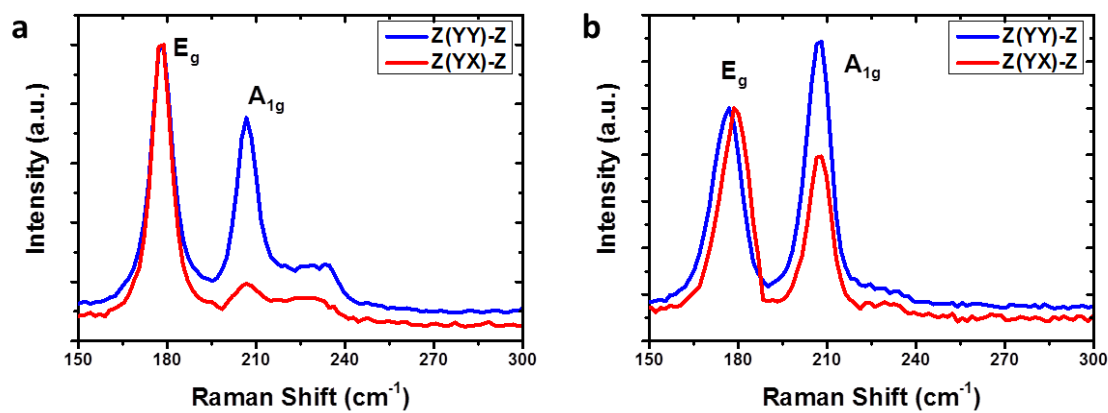

Figure 4 – Polarization response in Raman spectra of PtSe$_2$ films taken using the blue (488 nm) wavelength laser excitation (a) 1 nm thickness spectra (b) 5 nm thickness spectra

Figure 5(a) shows Raman spectra of PtSe$_2$ layers of a variety of thicknesses in the 100-600 cm$^{-1}$ region. The characteristic Si peak, observed at ~520 cm$^{-1}$, decreases in relative intensity as the PtSe$_2$ layer increases in thickness. When normalized to the $E_g$ peak at ~176 cm$^{-1}$, as highlighted in Figure 5(b), the intensity of the $A_{1g}$ peak at ~205 cm$^{-1}$ shows a clear change in relative intensity with increasing thickness from the thinnest to the thickest film, with the thicker films showing a dramatic increase in relative intensity in comparison to thinner films. As the intensity dependence on thickness is a clear indication of a Raman mode dependent on the out-of-plane motion of chalcogen atoms, this serves as further confirmation of the assignment of the $A_{1g}$ mode. A *LO* peak attributed to an overlap of $A_{2u}$ and $E_u$ is also observable at ~235 cm$^{-1}$, as discussed above, which consistently decreases in relative intensity with increasing film thickness. This is supported by the Raman analysis of commercially-available, bulk PtSe$_2$ crystals, grown by chemical vapor transport (CVT), in Figure S.6 in the supplementary data, where this peak is effectively absent for bulk samples. This Raman analysis of pristine PtSe$_2$ crystals also serves as confirmation of the high quality of the PtSe$_2$ synthesized in this work. To further analyze the relationship between the $A_{1g}$ mode and the thickness, Figure 5(c) shows the extracted thickness/intensity ratio of the $A_{1g}$ mode in PtSe$_2$, showing a dependence similar to that previously observed in MoS$_2$ and WS$_2$ thin films with increasing thickness[25, 26], and reflecting the increased van der Waals interactions between the layers with high layer numbers. Figure 5(d) shows rescaled spectra highlighting the red shift of the $E_g$ mode with increasing thickness, similar to the red shift observed in the $E^1_{2g}$ mode in TMDs such as MoS$_2$ with increasing layer number. This indicates that PtSe$_2$ in a 1T structure behaves similarly to other layered materials in 2H structures, where stacking-induced structural changes and long-range Coulombic interactions dominate the changing atomic vibration[25]. The red-shift of the $E_g$ mode with increasing thickness is also predicted by theoretical calculations. Neglecting the monolayer case, which does not match with our experimental realizations, the bilayer and bulk spectra can be compared, as shown in Figure 2 (d) and (e). While the $A_{1g}$ mode remains pinned at 193 cm$^{-1}$, the $E_g$ mode shifts from 166 cm$^{-1}$ to 155 cm$^{-1}$ as we compare the bilayer to the bulk PtSe$_2$. This energy shift of ~10 cm$^{-1}$ is in agreement with the shift observed by Raman scattering. Experimentally we observe a small red-shift in the $A_{1g}$ mode with increasing thickness but this is less pronounced than the shift in the $E_g$ mode.

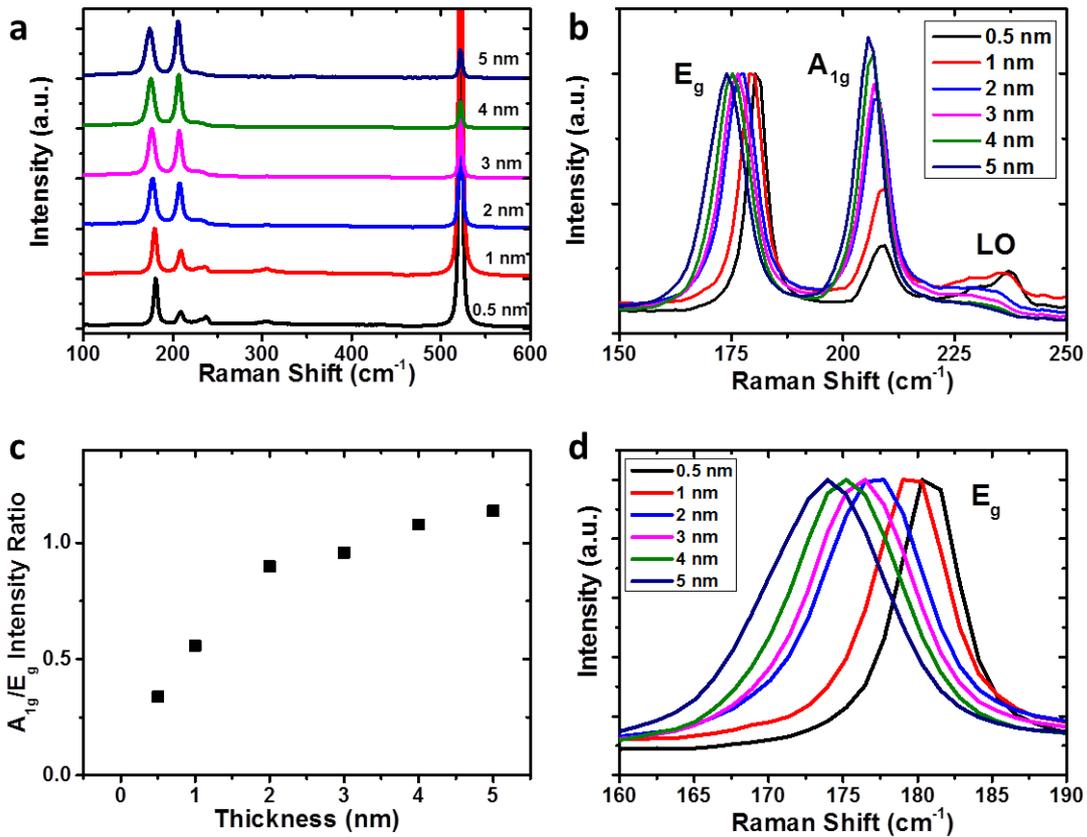

Figure 5 – Thickness dependent Raman spectra of $PtSe_2$ films taken using a 532 nm laser (a) Raman spectra of $PtSe_2$ films from 100 to 600 $cm^{-1}$. (b) Raman spectra of $PtSe_2$ films in the 150 to 250 $cm^{-1}$ region (c) Analysis of thickness dependent behavior of the $A_{1g}$ mode with starting Pt thickness (d) Raman Spectra from 160 to 190 $cm^{-1}$ highlighting the red shift of the $E_g$ mode with increasing thickness.

It should be noted that a general increase in signal intensity was observed for both modes with increasing film thickness, which is to be expected where there is more material for incoming light to interact with. An increase in the linewidth of the $E_g$ mode as the film thickness increases is also evident and merits further investigation. Similar broadening has been observed for the $E^1_{2g}$ mode of disordered $MoS_2$[27] and the effect observed here could be due to increased disorder with increasing layer thickness. This is supported by the broadening of the $E_g$ mode observed at edge sites of CVT crystals, as shown in Figure S.6 in the supplementary data. Additional Raman spectra acquired using a 633 nm excitation laser, which further highlight trends in spectral characteristics with changing thickness, are presented in the supplementary data, Figure S.7.

**Conclusions**

PtSe$_2$ thin films of controllable thickness have been synthesized by direct selenization of predeposited Pt layers. STEM studies confirm that the crystal lattice of is a 1T crystal structure, allowing the Raman spectrum of PtSe$_2$ to be analyzed and explained. A detailed analysis of the Raman-active modes in PtSe$_2$ has been presented. The $A_{1g}$ and $E_g$ out-of-plane and in-plane modes, respectively, have been identified using a variety of different laser wavelengths, and these assignments have been confirmed using polarization-dependent measurements and theoretical modelling. Furthermore, the thickness dependence of these modes has been investigated. This study will expedite further research into the synthesis and characterization of exotic TMDs.

**Acknowledgements**

N.M. acknowledges SFI for 14/TIDA/2329. M.O. acknowledges an Irish Research Council scholarship via the Enterprise Partnership Scheme, Project 201517, Award 12508. This work was supported by the SFI under contract no. 12/RC/2278 and PI 10/IN.1/I3030 and by the EU under contract n°604391 Graphene Flagship. SS and CM acknowledge the European Research Council for financial support (QUEST project). J.-Y.Z. acknowledges SFI for 14/IF/2499. J.K., J.M., and K.E. acknowledge support from the FWF Project No. P25721-N20 and the European Research Council (ERC) Project No. 336453-PICOMAT. T.P. acknowledges funding from the European

Union's Horizon 2020 research and innovation programme under the Marie Skłodowska-Curie grant agreement No. 655760 – DIGIPHASE.

**References**

1.  K. F. Mak, C. Lee, J. Hone, J. Shan and T. F. Heinz, *Physical Review Letters*, 2010, **105**, 136805.

2.  B. Radisavljevic, A. Radenovic, J. Brivio, V. Giacometti and A. Kis, *Nature Nanotechnology*, 2011, **6**, 147-150.

3.  K. S. Novoselov, A. K. Geim, S. V. Morozov, D. Jiang, Y. Zhang, S. V. Dubonos, I. V. Grigorieva and A. A. Firsov, *Science*, 2004, **306**, 666-669.

4.  V. Nicolosi, M. Chhowalla, M. G. Kanatzidis, M. S. Strano and J. N. Coleman, *Science*, 2013, **340**.

5.  J. N. Coleman, M. Lotya, A. O'Neill, S. D. Bergin, P. J. King, U. Khan, K. Young, A. Gaucher, S. De, R. J. Smith, I. V. Shvets, S. K. Arora, G. Stanton, H.-Y. Kim, K. Lee, G. T. Kim, G. S. Duesberg, T. Hallam, J. J. Boland, J. J. Wang, J. F. Donegan, J. C. Grunlan, G. Moriarty, A. Shmeliov, R. J. Nicholls, J. M. Perkins, E. M. Grieveson, K. Theuwissen, D. W. McComb, P. D. Nellist and V. Nicolosi, *Science*, 2011, **331**, 568-571.

6.  N. Scheuschner, R. Gillen, M. Staiger and J. Maultzsch, *Physical Review B*, 2015, **91**, 235409.

7.  A. C. Ferrari and D. M. Basko, *Nat Nano*, 2013, **8**, 235-246.

8.  X. Zhang, X.-F. Qiao, W. Shi, J.-B. Wu, D.-S. Jiang and P.-H. Tan, *Chemical Society Reviews*, 2015, **44**, 2757-2785.

9.  Y. Wang, L. Li, W. Yao, S. Song, J. T. Sun, J. Pan, X. Ren, C. Li, E. Okunishi, Y.-Q. Wang, E. Wang, Y. Shao, Y. Y. Zhang, H.-t. Yang, E. F. Schwier, H. Iwasawa, K. Shimada, M. Taniguchi, Z. Cheng, S. Zhou, S. Du, S. J. Pennycook, S. T. Pantelides and H.-J. Gao, *Nano Letters*, 2015, **15**, 4013-4018.

10. K. Ullah, S. Ye, Z. Lei, K.-Y. Cho and W.-C. Oh, *Catalysis Science & Technology*, 2015, **5**, 184-198.

11. H. L. Zhuang and R. G. Hennig, *The Journal of Physical Chemistry C*, 2013, **117**, 20440-20445.

12. R. Gatensby, N. McEvoy, K. Lee, T. Hallam, N. C. Berner, E. Rezvani, S. Winters, M. O'Brien and G. S. Duesberg, *Applied Surface Science*, 2014, **297**, 139-146.

13. C. Yim, M. O'Brien, N. McEvoy, S. Riazimehr, H. Schäfer-Eberwein, A. Bablich, R. Pawar, G. Iannaccone, C. Downing, G. Fiori, M. C. Lemme and G. S. Duesberg, *Scientific Reports*, 2014, **4**.

14. Y. Zhan, Z. Liu, S. Najmaei, P. M. Ajayan and J. Lou, *Small*, 2012, **8**, 966-971.

15. V. Blum, R. Gehrke, F. Hanke, P. Havu, V. Havu, X. Ren, K. Reuter and M. Scheffler, *Computer Physics Communications*, 2009, **180**, 2175-2196.

16. A. Tkatchenko and M. Scheffler, *Physical Review Letters*, 2009, **102**, 073005.

17. A. Togo, F. Oba and I. Tanaka, *Physical Review B*, 2008, **78**, 134106.

18. H. Nolan, N. McEvoy, M. O'Brien, N. C. Berner, C. Yim, T. Hallam, A. R. McDonald and G. S. Duesberg, *Nanoscale*, 2014, **6**, 8185-8191.

19. H. Wang, Z. Lu, S. Xu, D. Kong, J. J. Cha, G. Zheng, P.-C. Hsu, K. Yan, D. Bradshaw, F. B. Prinz and Y. Cui, *Proceedings of the National Academy of Sciences*, 2013, **110**, 19701-19706.

20. Y. Jung, J. Shen, Y. Liu, J. M. Woods, Y. Sun and J. J. Cha, *Nano Letters*, 2014, **14**, 6842-6849.

21. C. T. Koch, *Ph.D. Thesis*, 2002.

22. A. Molina-Sanchez and L. Wirtz, *Physical Review B*, 2011, **84**, 155413.

23. C. A. Arguello, D. L. Rousseau and S. P. S. Porto, *Physical Review*, 1969, **181**, 1351-1363.

24. D. Kong, H. Wang, J. J. Cha, M. Pasta, K. J. Koski, J. Yao and Y. Cui, *Nano Letters*, 2013, **13**, 1341-1347.

25. C. Lee, H. Yan, L. E. Brus, T. F. Heinz, J. Hone and S. Ryu, *ACS Nano*, 2010, **4**, 2695-2700.

26. A. Berkdemir, H. R. Gutiérrez, A. R. Botello-Méndez, N. Perea-López, A. L. Elías, C.-I. Chia, B. Wang, V. H. Crespi, F. López-Urías and J.-C. Charlier, *Scientific Reports*, 2013, **3**, 1755.

27. S. Mignuzzi, A. J. Pollard, N. Bonini, B. Brennan, I. S. Gilmore, M. A. Pimenta, D. Richards and D. Roy, *Physical Review B*, 2015, **91**, 195411.

# Supplementary Data – Raman Characterization of Platinum Diselenide Thin Films

Maria O'Brien[a,b†], Niall McEvoy[a,b†*], Carlo Motta[a,c], Jian-Yao Zheng[a,c], Nina C. Berner[a,b], Jani Kotakoski[d], Kenan Elibol[d], Timothy J. Pennycook[d], Jannik C. Meyer[d], Chanyoung Yim[a,b], Mohamed Abid[e], Toby Hallam[a,c], John F. Donegan[a,c], Stefano Sanvito[a,c] and Georg S. Duesberg[a,b*]

[a] Centre for the Research on Adaptive Nanostructures and Nanodevices (CRANN) and Advanced Materials and BioEngineering Research Institute (AMBER), Trinity College Dublin, Dublin 2, Dublin, Ireland

[b] School of Chemistry, Trinity College Dublin, Dublin 2, Dublin, Ireland

[c] School of Physics, Trinity College Dublin, Dublin 2, Dublin, Ireland

[d] Faculty of Physics, University of Vienna, Boltzmanngasse 5, A-1090 Vienna, Austria

[e] KSU-Aramco Center, King Saud University, Riyadh 11451, Saudi Arabia

†These authors contributed equally. *E-mail: nmcevoy@tcd.ie and duesberg@tcd.ie

## S1 - Furnace Schematic

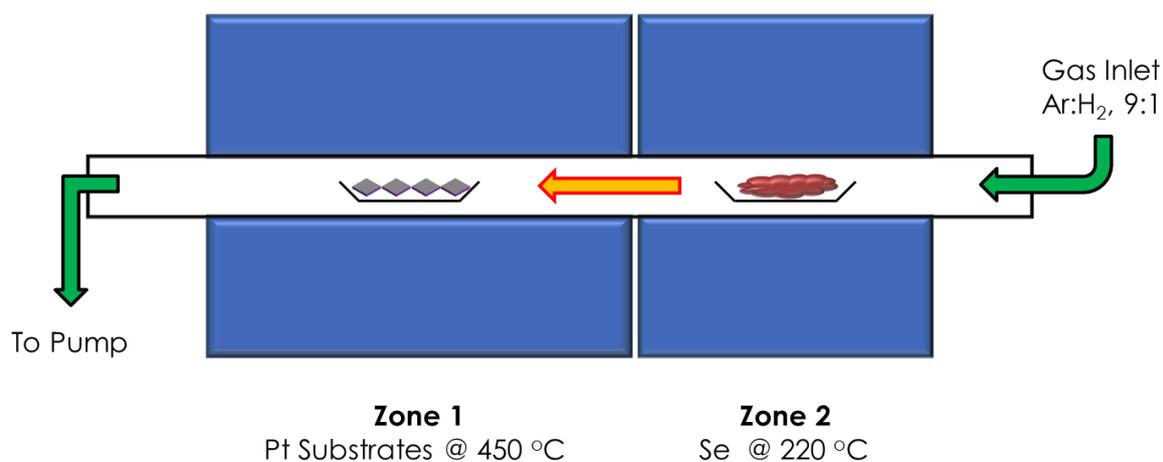

Figure S.1 – A schematic of the furnace setup described in the experimental procedures for fabrication of the PtSe$_2$ thin films.

## S2 – Image simulation of PtSe$_2$ Lattice

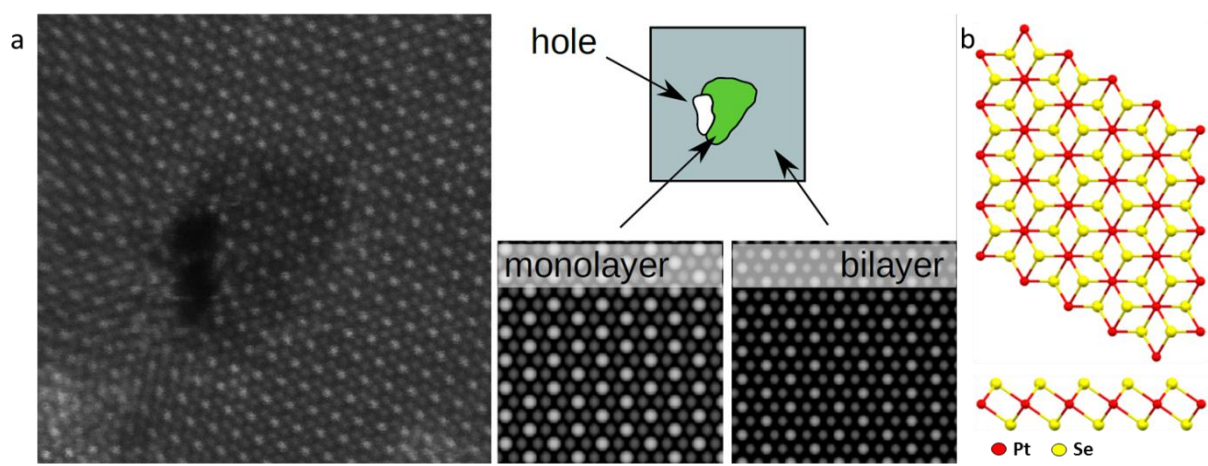

Figure S.2 – (a) An area of the PtSe$_2$ membrane showing a hole, presumably caused by the imaging electrons during the HAADF-STEM imaging, and a monolayer area within a larger bilayer structure. The interpretation of the layer thicknesses is confirmed by image simulations with QSTEM software[1] (right) for mono- and bilayer structures of 1T phase of PtSe$_2$. (b) A schematic of the 1T crystal structure of PtSe$_2$.

## S3 – Additional XPS Spectra

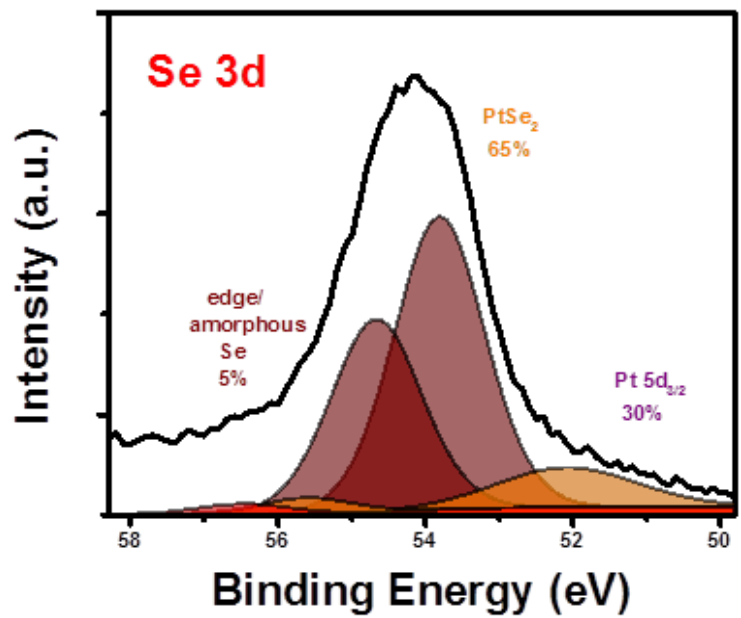

Figure S.3 – XPS spectrum of 0.5 nm PtSe$_2$ thin film showing the Se 3d peak.

## S4 - Electronic Band Structures and Expanded Phonon Dispersion Curves

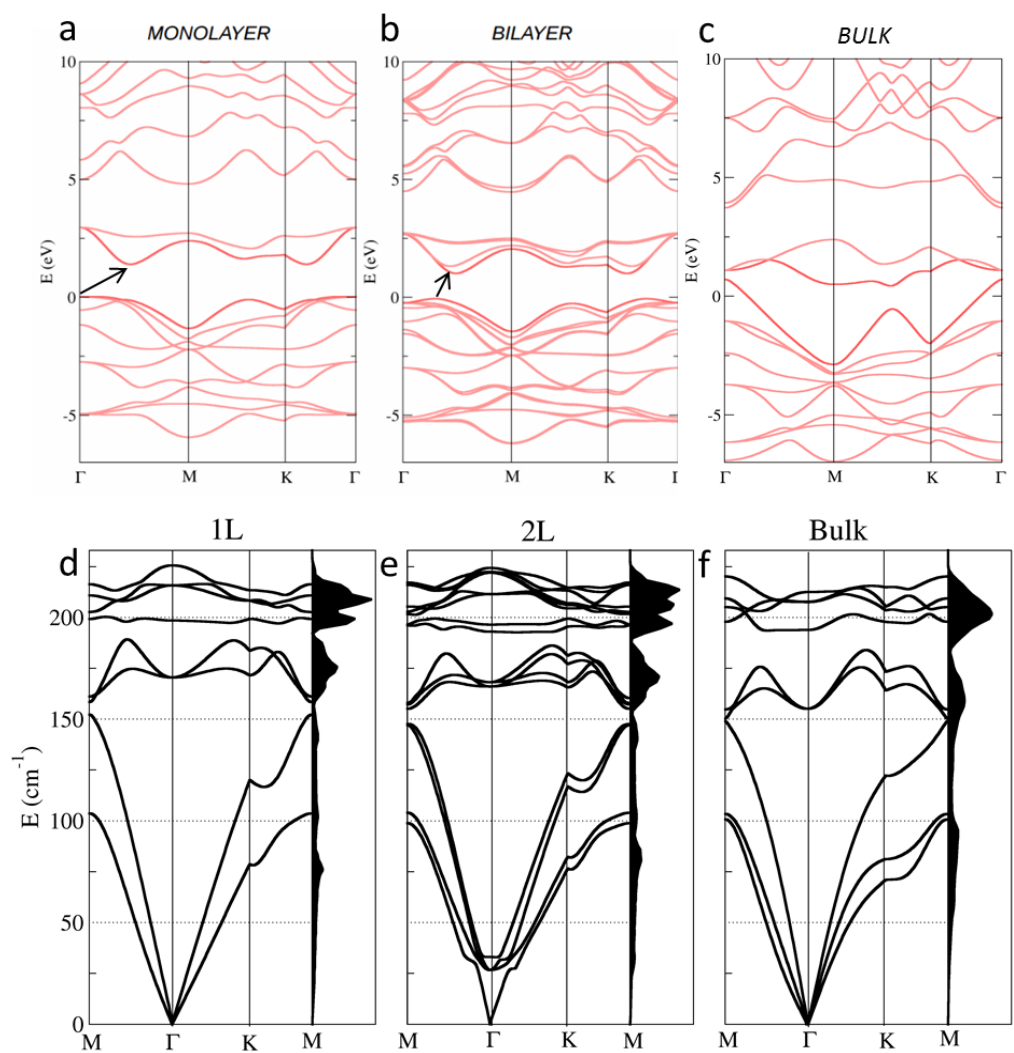

Figure S.4 – Calculated electronic bandstructures of (a) monolayer (b) bilayer and (c) bulk PtSe$_2$. Expanded phonon dispersion curves from Figure 2 in the main text are shown for (d) monolayer (e) bilayer and (f) bulk PtSe$_2$.

| Mode Prediction | LDA | GGA |
| --- | --- | --- |
| $E_g$ | 183.11093 | 170.48864 |
| $E_g$ | 183.15949 | 170.49728 |
| $A_{1g}$ | 200.78692 | 198.51035 |
| $E_u$ | 233.23271 | 215.76595 |
| $E_u$ | 233.40321 | 215.78960 |
| $A_{2u}$ | 237.92582 | 225.62144 |

Table S.1 – Extracted vibrational frequencies for LDA and GGA calculation methods.

## S5 – Additional Calculations of PtSe$_2$ Bandstructure with Spin-Orbit Interactions

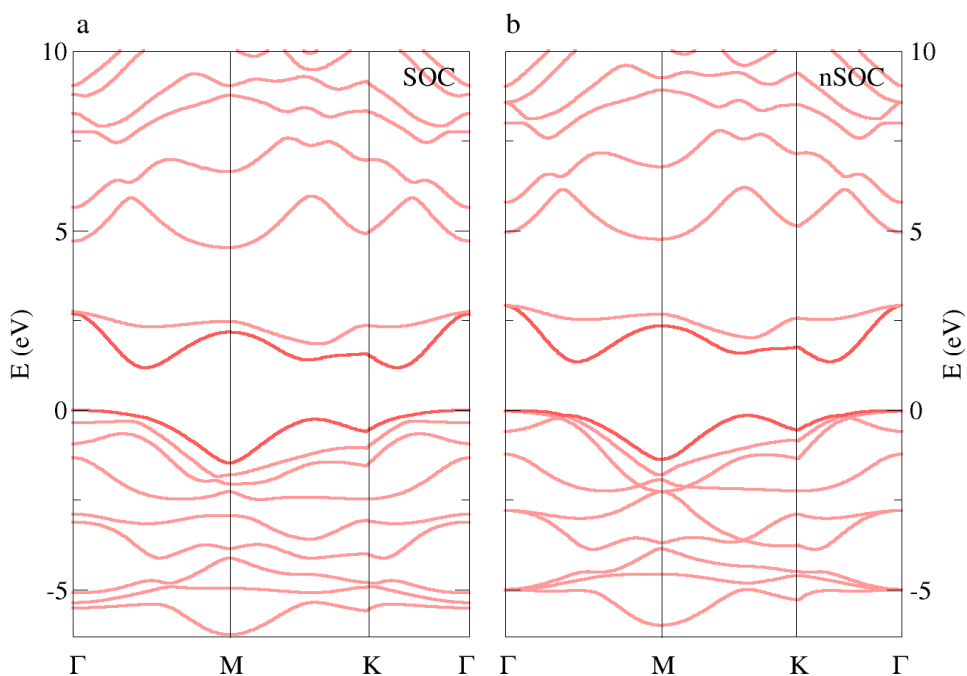

Figure S.5 Electronic bandstructure of PtSe$_2$ calculated with (a) and without (b) spin orbit coupling. The valence and conduction bands are highlighted with a darker color.

## S6 - Raman Analysis of PtSe$_2$ Crystal Grown by Chemical Vapor Transport

A bulk PtSe$_2$ crystal, grown by chemical vapor transport (CVT), was purchased from 2D Semiconductors and used as a reference sample for Raman analysis. Optical micrographs of this crystal are shown in Figure S.6(a, b) in which edge and basal planes are evident. Much like the films grown by vapor-phase selenization of predeposited Pt, the CVT crystal displays clear $E_g$ and $A_{1g}$ modes in its Raman spectrum. Analysis of this crystal allowed for observations to be made on the differences in Raman spectra between edge and basal plane sites. Representative Raman spectra, normalized to the $E_g$ mode intensity, extracted from these edge and basal plane sites are shown in Figure S.6(c). The spectrum of a basal plane site is similar to that of the thickest films examined in the main text, albeit with a lower relative $A_{1g}$ mode intensity, and can be considered representative of bulk-like PtSe$_2$. The intensity of both Raman modes is greater in the edge sites, as shown in the maps in Figure S.6(d, e), which are taken over the area marked by the red box in Figure S.6(b). This is probably due to the greater sample interaction volume in these areas due to the exposed edges. Interestingly, the relative intensity of the $A_{1g}$ mode is considerably greater in the edge site regions, as shown in the map in Figure S.6(f). This is most likely due to the greater contribution from out-of-plane modes in this region. Similar trends have been seen previously for crystals of MoS$_2$[2]. The high relative intensity of the $A_{1g}$ mode in the Raman spectra of thicker films presented in Figures 3, 4 & 5 of the main text can be attributed to the polycrystalline nature of the samples. A broadening of the $E_g$ mode is also observed at edge sites. Similar broadening is observed with increasing film thickness in Figure 5(d) of the main text.

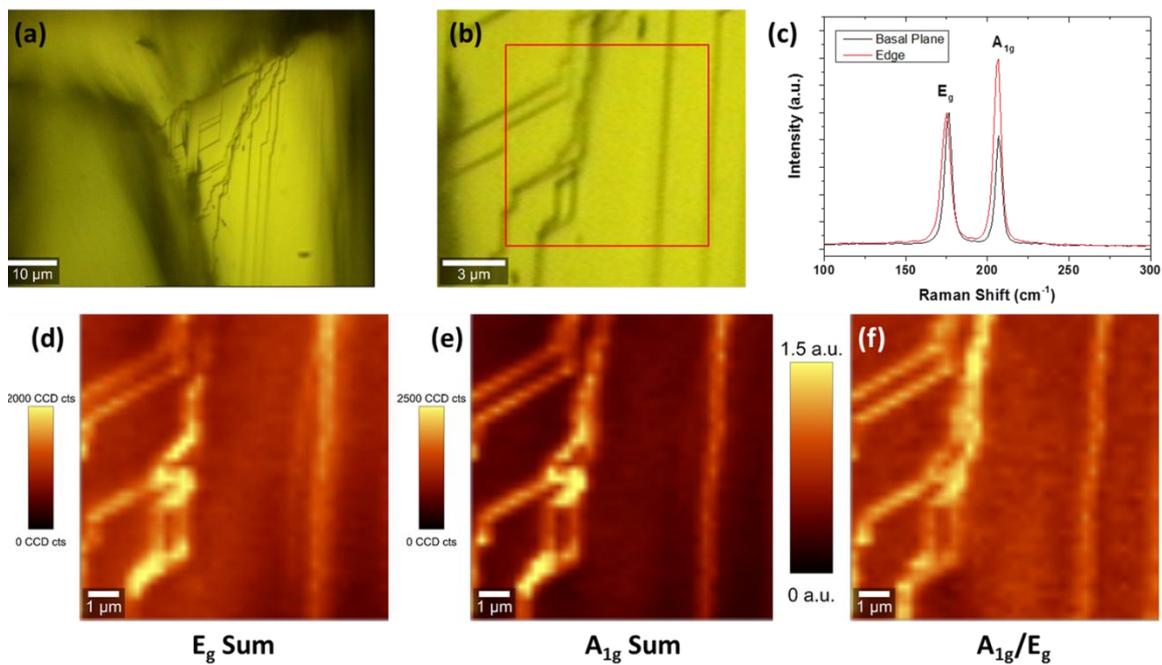

Figure S.6 (a, b) Optical micrographs of CVT-grown PtSe$_2$ crystal. (c) Raman spectra extracted from basal plane and edge regions. (d, e) Peak sum Raman maps from area marked by red box in (b). (f) $A_{1g}/E_g$ ratio map from area marked in (b)

**S7 – Thickness Dependent Raman Studies with 633 nm Excitation Laser**

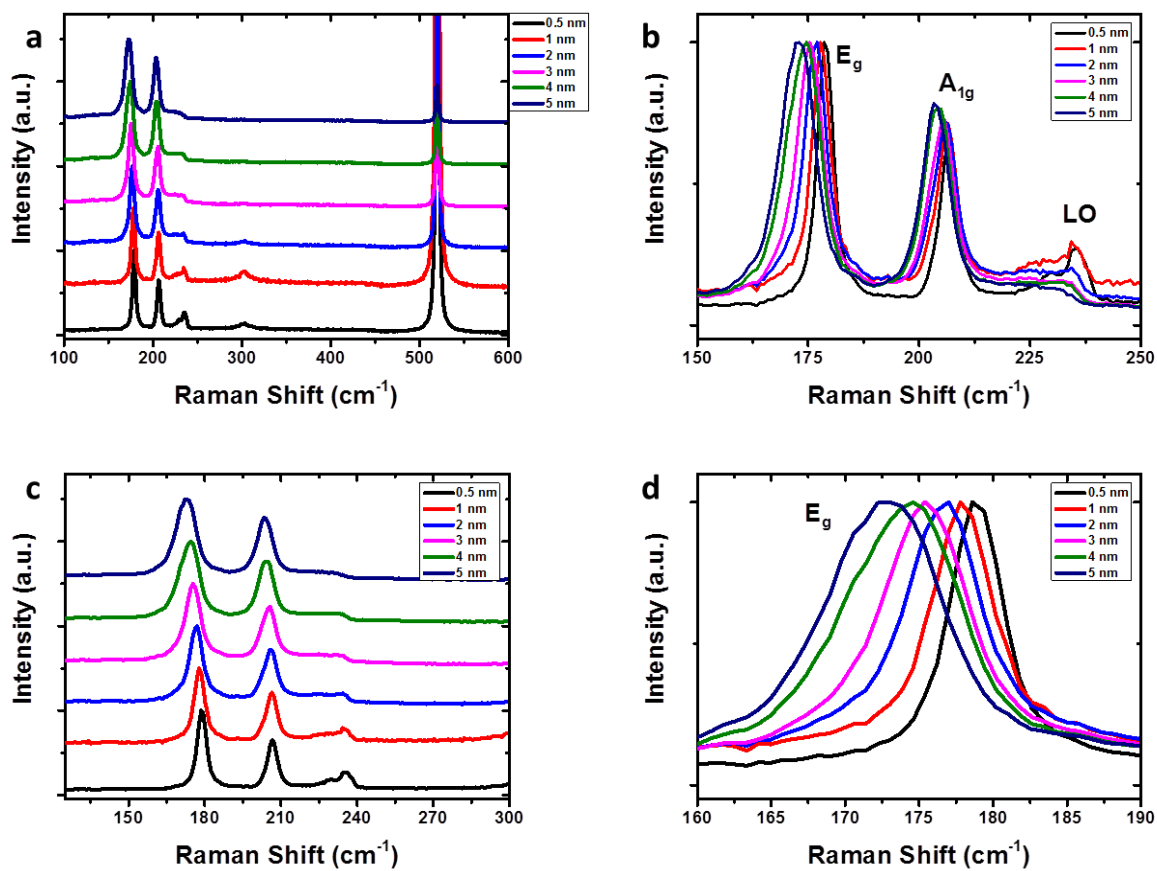

Figure S.7 (a-d) Raman spectra of the same samples probed in Figure 5 of the main text, acquired using a 633 nm excitation laser.

**References**

1. C. T. Koch, *Ph.D. Thesis*, 2002.
2. S. M. Tan, A. Ambrosi, Z. Sofer, Š. Huber, D. Sedmidubský and M. Pumera, *Chemistry – A European Journal*, 2015, **21**, 7170-7178.